\documentclass[aps, pre,floats,superscriptaddress,floatfix,twocolumn,10pt]{revtex4-1}
\usepackage{amssymb,amsmath}
\usepackage{color,soul}
\usepackage{graphicx}
\usepackage{hyperref}
\begin{document}
\date{\today}
\title{Universality classes of absorbing phase transitions in generic branching-annihilating particle systems}
\author{Bijoy Daga}\email{bijoydaga@imsc.res.in}
\affiliation{The Institute of Mathematical Sciences, C.I.T Campus, Taramani, Chennai-600113, India}
\author{Purusattam Ray}\email{ray@imsc.res.in}
\affiliation{The Institute of Mathematical Sciences, C.I.T Campus, Taramani, Chennai-600113, India}
\affiliation{Homi Bhabha National Institute, Training School Complex, Anushakti Nagar, Mumbai-400094, India}
\begin{abstract}
We study absorbing phase transitions in systems of branching annihilating random walkers 
and pair contact process with diffusion on a one dimensional ring, where the walkers 
hop to their nearest neighbor with a bias 
$\epsilon$. For $\epsilon=0$, three 
universality classes: 
directed percolation (DP), parity conserving (PC) and pair contact process 
with diffusion (PCPD) are typically observed in such systems. We find 
that the introduction of $\epsilon$ does not change the DP 
universality class but alters the other two universality classes. For non-zero $\epsilon$, the PCPD class crosses over to DP and the PC class changes to a new universality class.  
\end{abstract}

\maketitle

\section{Introduction}
Many reaction-diffusion systems  show 
a second-order phase transition from 
a fluctuating active state to 
a non-fluctuating absorbing state as some control
parameter is tuned \cite{hinrichsen_rev, odor_rev, hinrichsen_book}. A wide range of models 
corresponding to  phenomena such as epidemic spreading \cite{mollison_epidemic}, 
catalytic 
chemical reactions \cite{chemical}, transport in disordered 
media \cite{havlin_transport}, forest fire \cite{albano_fire}, biological 
evolution \cite{lipowski_evolution}, surface
roughening \cite{hinrichsen_rough, roughening}, self-activated biological 
structures \cite{berry_biology}, etc
show absorbing phase transitions.
These transitions are classic examples of nonequilibrium phase transitions. 
Studying the critical behavior and universality classes of  such transitions
is extremely important from theoretical perspective and for understanding the phase transition in reaction diffusion systems.
\par A large number of absorbing phase transitions in nonequilibrium
systems have been observed to belong to the directed percolation (DP) universality class.
It has been conjectured by Janssen and Grassberger\cite{DP_conjecture} that  continuous absorbing phase transitions in reaction-diffusion system with short-range interactions, characterized by a non-negative scalar order parameter and with no additional symmetries, conservation laws and quenched randomness  belong generically to the DP universality class. The robustness of DP universality class has been a matter
of great interest among researchers. The parity conserving (PC) universality class \cite{pc_class}, the universality class of pair contact process 
with diffusion \cite{pcpd_class}, the voter universality class \cite{voter_class}
and the Manna universality class in sandpile models \cite{manna_class} are some 
noteworthy
universality classes in nonequilibrium phase transitions whose critical behavior is different from that of DP.  
\par In this work, we have focused on branching annihilating 
random walks (BARW) and pair contact process with 
diffusion (PCPD), where three  universality classes, namely DP, PC and PCPD 
have been reported. In BARW, 
a diffusing random walker $A$ can branch to 
produce $m~(m>0)$ new off-springs $A \to (m+1)A$, or two of them  
can annihilate ($2A \to \O$)  upon contact.
The parity of the system which is defined as 
the total number of walkers modulo 2 is not conserved when $m$ is odd. Depending 
upon parity, the critical behavior in such systems  is DP when $m$ is odd 
and PC when $m$ is even \cite{pc_class}. PC class 
is also referred to as directed ising (DI) class because PC critical behavior can 
also be realized in spin systems when a spin-flip Glauber dynamics compete 
with spin-exchange Kawasaki dynamics \cite{menyhard_nikem}.  In PCPD \cite{pcpd_class}, two 
diffusing walkers
in contact only can produce new off-springs ($2A \to 3A$) or they can 
annihilate ($2A \to \O$). Unlike in PC, parity 
does not affect the critical behavior in PCPD\cite{parity_PCPD}.
\par The crossover behavior between these universality classes have been extensively
studied. It has been found that the PC class crosses over to DP  by introducing 
a dynamics which breaks the modulo 2 conservation. 
This is done by producing 
both even and odd number of off-springs while branching \cite{park_park_PC_DP}. 
In addition to  PCPD dynamics, if unary branching and annihilation are introduced, 
then PCPD crosses over to DP  when the 
unary process does not conserve parity \cite{pcpd_dp, pcpd2_dp} or to PC if it conserves parity \cite{pcpd2_dp}. This suggests that parity and  $n-$narity of the branching 
process plays a crucial 
role in determining the universality class of the transition. It is also found 
that  diffusion,  or its absence affects the critical behavior
in a major way. For example, in absence of diffusion, parity 
conserving BARW models having spatially 
asymmetric branching can have additional conservation laws depending upon the 
initial conditions, 
and consequently, the decay exponent varies from 
the PC exponent \cite{hinrichsen_rough}.
The PCPD class crosses over continuously 
to DP when solitary 
diffusing walkers are annihilated upon contact with a certain probability which 
determines the value of the critical exponents \cite{noh_park}. 
When  diffusion of single walkers is completely forbidden in PCPD, although the system has 
multiple absorbing states, its critical behavior switches over to
DP \cite{pcp_jensen}. BARW has 
also been studied with Le\'vy walkers.  
The  additional long-range correlations  that builds up in the system due 
to long-range
interactions via Le\'vy flights  leads to 
continuous variation 
of critical exponents for both DP and PC universality classes  \cite{albano_DP_conti, vernon_PC_conti}. The above mentioned perturbations either change the parity, 
or brings in additional conservation laws, long-range interactions, or 
arrests the diffusion dynamics.
\par We ask what happens to DP, PC and PCPD universal critical behaviors
when only local  perturbations are introduced to the underlying diffusion dynamics without
affecting parity of the system, creating any long-range interactions, or bringing
in additional conservation laws. Specifically, 
we study  BARW and PCPD 
on a one dimensional periodic chain where the walkers hop to 
their nearest neighbor with a bias 
$\epsilon$. A walker at a given site  diffuses towards its 
nearest neighbor 
with probability  $\frac{1}{2}+\epsilon~(0\leq \epsilon \leq 1/2)$ and in the opposite 
direction with the complimentary probability $\frac{1}{2}-\epsilon$. For 
$\epsilon=0$, the walker performs a simple  random walk and for $\epsilon=1/2$, the  walker moves ballistically towards its nearest neighbor.
It is to be noted that the  bias on a walker at different sites are 
uncorrelated and so is  the bias on the walker at different times. The process retains its Markovian nature and unlike the problem with Le\'vy walker, there is 
no additional long-range interaction that is present in the system. However,  
the bias  hinder 
the diffusion of walkers away from their parent cluster. Thus, for $\epsilon>0$, branching 
and annihilation within a cluster become the dominant processes.  
The case of annihilating random walkers ($i.e.$ no branching) in presence of the
bias $\epsilon$ has been studied before \cite{sen_ray} and it was found  that the  
decay exponent changes from a value 1/2 without bias to 1 when bias 
is introduced. This suggests that under the bias, the random 
walkers at large times behave as ballistic walkers.
When branching is turned on, there  is an absorbing  phase 
transition for $\epsilon=0$ \cite{pc_class, pcpd_class}. With $\epsilon>0$,
one would expect that  
the transition between absorbing and active phases  to occur at higher branching rates because of the enhanced annihilation.
An important question to ask is whether this bias will affect the critical 
behavior of the transition and what are the possible universality classes 
it can give rise to. 
We study the problem using Monte Carlo simulations. We find that  
non-zero $\epsilon$ retains the universality class of  DP, where as changes the PCPD class to DP and the PC class changes to a new universality class.

\section{Models}
On a one dimensional lattice, the BARW with nearest 
neighbor bias $\epsilon$ ($0\leq \epsilon \leq 1/2$) is defined in the following way: with probability $p$, a random walker $A$
diffuses and with the complimentary probability $1-p$ it branches to 
produce $m(m>0)$ 
off-springs: $A \to (m+1) A$ at its nearest neighboring sites.  When a 
walker diffuses, it does so
with a probability $\frac{1}{2}+\epsilon$ towards its nearest 
neighboring walker and with probability $\frac{1}{2}-\epsilon$ 
in the opposite direction. 
When two walkers meet at the same site, they annihilate $(2A \to \O)$. 
For large values of  $p$, all walkers get annihilated and the system goes to 
an absorbed state. When $p$ is small, the branching rate being higher, the system 
has a finite density of walkers even at large times and hence the system remains active forever. Therefore, by varying $p$, one can observe absorbing phase transitions in such systems at a particular critical value of $p=p_c$. When $m$ is even, the number 
of walkers modulo 2 is conserved at all times. This symmetry is called the parity. For $\epsilon=0$, the critical behavior of the transition between 
active and absorbing states depends upon parity. The critical 
behavior in these systems  belong to PC when $m$ is even 
and to DP when $m$ is odd \cite{pc_class}. We vary $\epsilon$ and 
find out how the critical behavior of the absorbing phase transitions change for 
$m=1$ and $m=4$ cases. 
\par We also study the effect of the nearest neighbor  
bias  on a binary process like PCPD \cite{pcpd_class}, where 
branching can occur only when two random walkers are placed side by side. 
A walker is selected at random and its neigboring site (left or right) is chosen with equal probability. The system  then evolves by  following one of the 
three processes with the respective assigned probabilities as described below: \\ 
(i) with probability $q(1-D)$, the walker and its neighbor in 
the chosen site are annihilated $(2A \to \O)$.\\
(ii) if the chosen neighboring site 
is occupied and the next nearest site in the direction of the neighboring  
site is empty, a new 
walker is created ($2A \to 3A$) at the next nearest site with 
probability $(1-q)(1-D)$. \\
(iii) a walker diffuses with probability $D$ to one of its neighboring site (left or right) if it is empty.  In this step the neighboring site 
is not chosen with equal probability. The target site for the diffusing walker is chosen 
with probability $1/2+\epsilon$ towards its nearest neigboring walker and with probability 
$1/2-\epsilon$ in the opposite direction. 
\par For $\epsilon=0$, the critical point and the critical exponents for absorbing phase transition in PCPD  generally depends upon both $q$ and $D$ \cite{pcpd_class}.
In this work, we find out the critical 
point $q_c$ for a fixed value of $D=1/2$ and study how the 
critical behavior changes as $\epsilon $ is varied. 

\begin{figure}
\includegraphics[height=6cm]{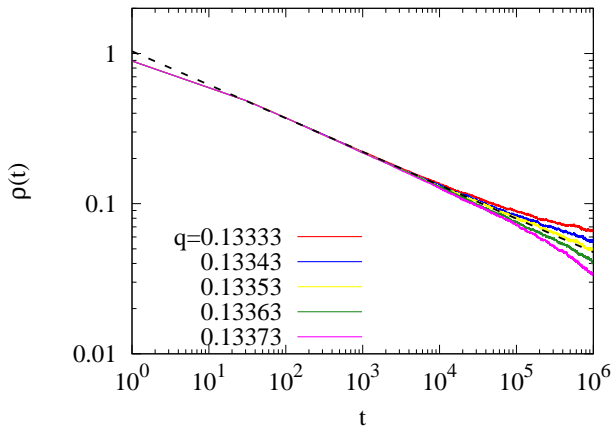}
\caption{Log-log plot of $\rho(t)$ vs. $t$ when $\epsilon=0$ in PCPD model for various values 
of the parameter $q$. For $q=0.13353$, the density decays with a power law as shown 
by the dotted line, thus giving an estimate of the critical point $q_c=0.13353$. The 
slope of the dotted line gives the estimate of the decay exponent $\alpha\sim0.221$. Here $L=25000$. }
\label{fig:alpha_eps0}
\end{figure}

\section{Simulation }
\begin{figure}
\includegraphics[height=6cm]{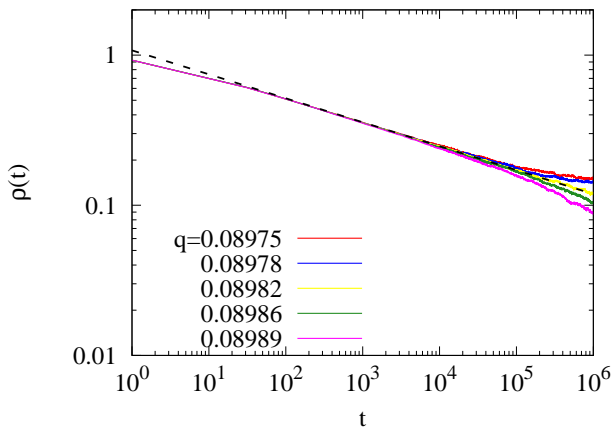}
\caption{Log-log plot of $\rho(t)$ vs. $t$ when $\epsilon=0.5$ in PCPD model for different values 
of the parameter $q$. For $q=0.08982$, the density decays with a power law as shown 
by the dotted line, thus giving an estimate of the critical point $q_c=0.08982$. The 
slope of the dotted line gives the estimate of the decay exponent $\alpha\sim0.159$. Here $L=25000$.}
\label{fig:alpha_eps0.5}
\end{figure}

To simulate BARW and PCPD, we start with a fully occupied 
lattice at time $t=0$ and measure the average density 
of walkers $\rho(t)=\langle s_i(t)\rangle$ as a function of time. Here, the 
$\langle \cdots \rangle$ represents average over configurations. 
The variable $s_i(t)$ takes the value $1$
when site $i$ is occupied by a  walker and $0$ 
when it is unoccupied.
As $t \to \infty$, $\rho(t)$ saturates to 
a positive value $\rho_a$, if the system is in the active
phase ($p<p_c$ or $q<q_c$) and decays to zero 
in the absorbing phase ($p>p_c$ or $q>q_c$). Thus, effectively, $\rho(t)$ acts as an order parameter
for the system. At the
critical point where $p=p_c$ for BARW and $q=q_c$ for PCPD, density decays  
with time as power law,
\begin{equation}
\rho(t) \sim t^{-\alpha}
\end{equation}
$\alpha$ being the decay  exponent. The critical points $p_c$, $q_c$ and 
the decay exponent $\alpha$ can be estimated by plotting $\rho(t)$ vs $t$ for 
different values of $p$ and $q$ respectively. At the critical point, a power law is obtained. 
For illustration, in Figures \ref{fig:alpha_eps0} and \ref{fig:alpha_eps0.5}, we  make an estimate of the critical point $q_c$ and the exponent $\alpha$ in the PCPD model for
$\epsilon=0$ and $\epsilon=0.5$ respectively. For different values of 
$\epsilon$, the estimate of the critical point and $\alpha$ for BARW and PCPD  has been compiled in Table \ref{table:exponents}. 

\par
\begin{figure}
\includegraphics[height=6cm]{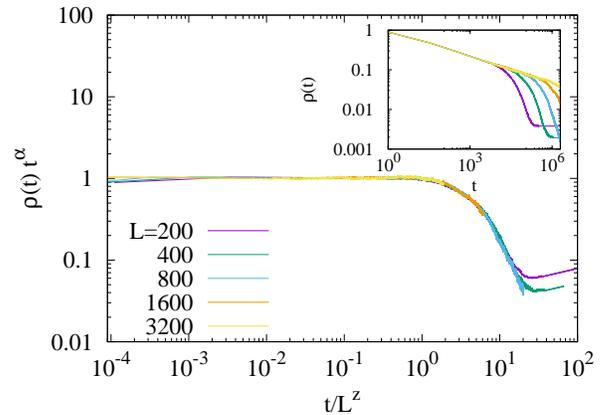}
\caption{Numerical estimate of the dynamical exponent $z$ in PCPD using finite size scaling: Log-log plot of $\rho(t)t^{\alpha}$
vs $t/L^z$ at $q_c=0.13353$ when $\epsilon=0$ for  L=200, 400,800,1600 and 3200. A good data collapse is obtained  for $z=1.72$. In the inset, the corresponding 
unscaled data has been plotted. }
\label{fig:z_eps0}
\end{figure}

\begin{figure}
\includegraphics[height=6cm]{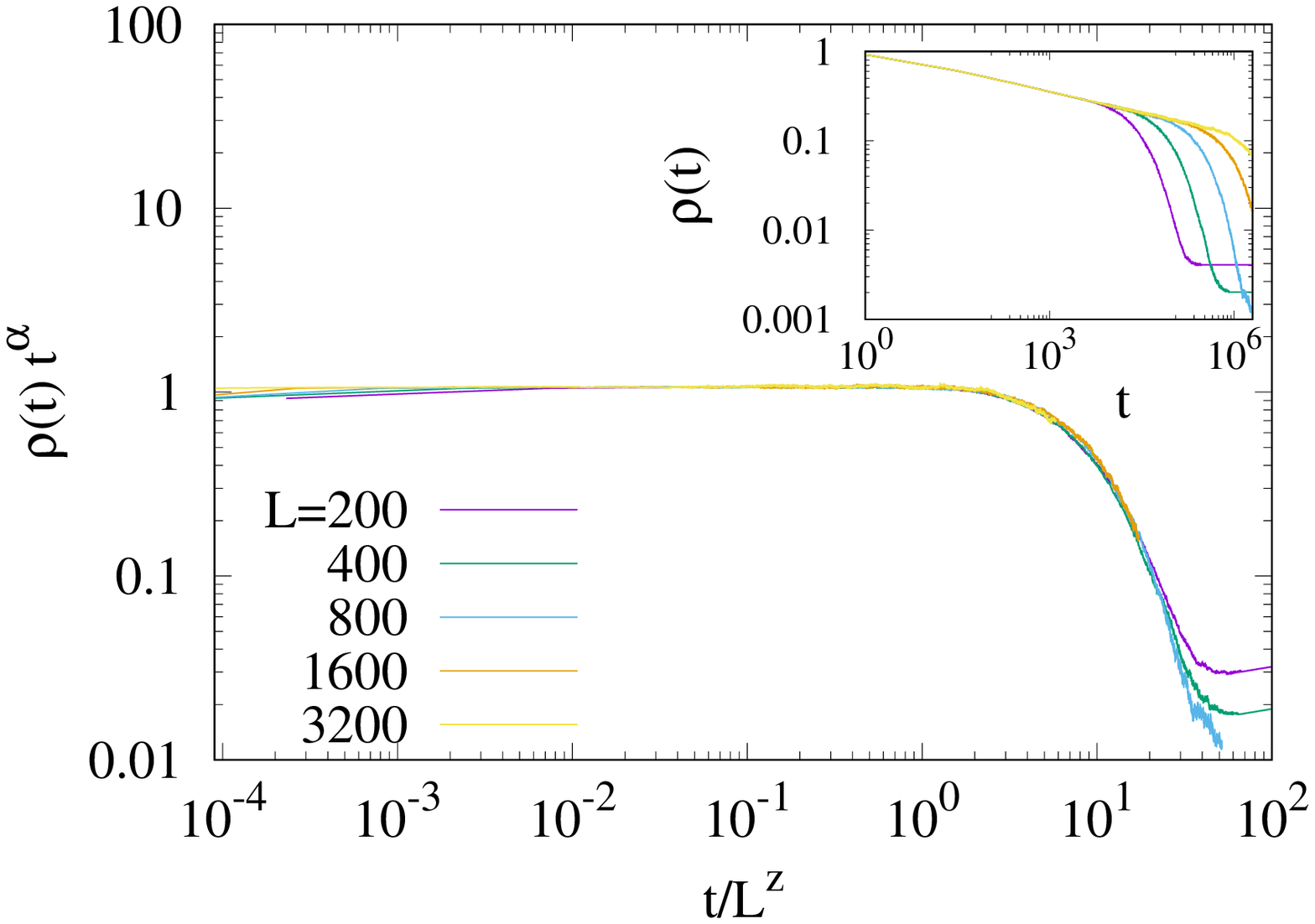}
\caption{Numerical estimate of the dynamical exponent $z$ in PCPD using finite size scaling: Log-log plot of $\rho(t)t^{\alpha}$
vs $t/L^z$ at $q_c=0.08982$ when $\epsilon=1/2$ for  L=200, 400, 800, 1600 and 3200. A good data collapse is obtained  for $z=1.58$. In the inset, the corresponding 
unscaled data has been plotted.}
\label{fig:z_eps0.5}
\end{figure}

 The dynamical  exponent $z$ can be determined from the finite size 
scaling analysis. In 
a finite system of size $L$, the decay of $\rho(t)$ as a function of $t$ at the critical point 
has the scaling form
\begin{equation}
\rho(t,L) \sim t^{-\alpha}f(t/L^z), 
\label{eq:z_exp}
\end{equation}
where $f$ is a scaling function. $f(x)$ is a constant 
for $x<1$ and decays exponentially for $x>1$.  Once $\alpha$ has been measured, one 
can then determine $z$ using Eq.~\eqref{eq:z_exp}. At the critical point, the 
curves $\rho(t)t^{\alpha}$ vs $t/L^z$ for different
values of $L$ collapse to a single curve. In 
Figures \ref{fig:z_eps0} and \ref{fig:z_eps0.5}, we use the finite 
size scaling method to 
measure $z$ for PCPD when  $\epsilon=0$ and $\epsilon$=0.5 respectively. 
The data for different values of $L$ collapses 
when $z=1.72$ for $\epsilon=0$ and $z=1.59$ for $\epsilon=0.5$  
\par
\begin{figure}
\includegraphics[height=6cm]{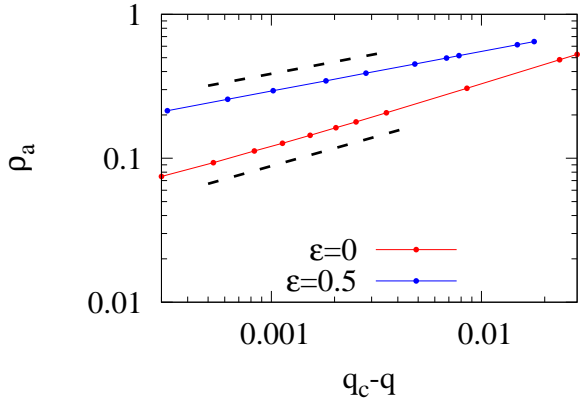}
\caption{Log-log plot of $\rho_a$ vs $q_c-q$ in PCPD for $\epsilon=0$ and $\epsilon=0.5$. 
The slopes of the lines (shown by the dotted lines) near $q \to q_c$ gives an  estimate of the exponent $\beta$. For $\epsilon=0$, $\beta\sim0.430$ and for $\epsilon=0.5$, $\beta\sim 0.275$. }
\label{fig:beta_pcpd}
\end{figure}

 The order parameter exponent $\beta$  characterises the algebraic decay of  the 
density $\rho_a$  as one approaches the 
critical point $q_c$ in the active phase $(q \to q_c^{-})$:  
\begin{equation}
\rho_a \sim (q_c-q)^{\beta}~
\end{equation}
Figure \ref{fig:beta_pcpd} shows the logarithmic plot of $\rho_a$ vs $q_c-q$ and  estimated values of $\beta$ in PCPD for $\epsilon=0$ and $\epsilon=0.5$.
\par
\begin{figure}
\centering\includegraphics[width=8cm]{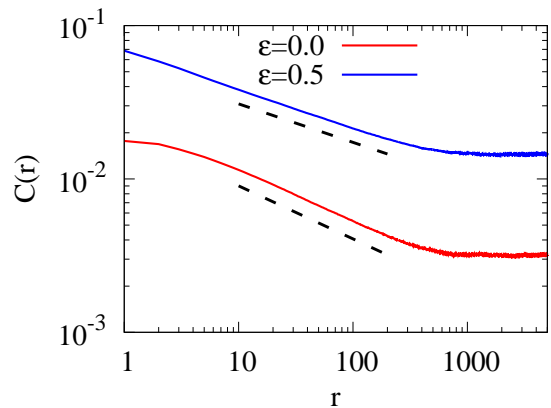}
\caption{Two point correlation function $C(r)$ measured at the critical point $q_c$ in PCPD
for $\epsilon=0$ and  $\epsilon=0.5$ at $t=10^5$ for a system of 
size $L=25 \times 10^3$. The slopes shown by dotted line gives the estimate 
of $\theta$. For $\epsilon=0$, $\theta \sim 0.346$ 
and $\theta \sim 0.251$ for $\epsilon=0.5$.}
\label{fig:2pt_pcpd}
\end{figure}

 We also measure the two-point spatial 
correlations, $C(r)=\langle s_i(t)s_{i+r}(t)\rangle$, where $\langle \cdots\rangle$ 
is the configuration average. At the critical point, $C(r)$
decays as a power law,
\begin{equation}
C(r) \sim r^{-\theta}
\end{equation}
where the exponent $\theta=z\alpha$. In Figure \ref{fig:2pt_pcpd}, we 
plot $C(r)$ vs. $r$ in PCPD for $\epsilon=0$ and $\epsilon=0.5$, measured 
at $t=10^5$. In Table \ref{table:exponents}, we put the  values 
of $\theta$ obtained from simulations for various values of   $\epsilon$. The exponents 
$\theta$, $\alpha$ and $z$ seems to satisfy the scaling relation $\theta=z \alpha$ for all
$\epsilon$.


\begin{table}[t!]
\centering
\begin{tabular}{|c|c|c|c|c|c|c|} 
\hline
Model & $\epsilon$ & $p_c$ & $\alpha$ & $\beta$  &   $z$ & $\theta$ \\ 
\hline
BARW & 0 &0.1070(1)& 0.161(1)~ &0.278(1) &1.58(1)&0.251(1)\\
m=1	 & 0.1&0.08355(3)&0.159(2) &0.276(1) &1.59(1) &0.250(1)\\
	& 0.3 &0.05819(4) &0.159(2) &0.275(2) &1.58(1)&0.251(2) \\
	& 0.5 &0.04469(2) & 0.158(1) &0.276(1) &1.58(1)&0.250(2) \\
\hline
BARW & 0 &0.7215(5) & 0.284(3)  & 0.92(4) & 1.75(1) & 0.491(2)\\ 
m=4	 & 0.1 &0.5620(2) & 0.232(4) & 0.55(1) & 1.72(1)&0.381(5) \\
	& 0.3 &0.4182(2) &0.229(3) &0.53(1) & 1.71(2)&0.378(3)\\
	& 0.5 &0.3369(2)& 0.224(5) & 0.51(2) & 1.74(2)&0.376(2) \\
\hline
     & & $q_c$ & & & &\\
\hline     
PCPD & 0 &0.13353(6)& 0.221(3)  & 0.43(1) & 1.72(2) & 0.346(1) \\ 
	 & 0.1 &0.11898(4)& 0.159(4) & 0.28(1) &1.59(2) & 0.254(2)\\
	& 0.3 &0.10087(3)& 0.159(1) & 0.275(3) & 1.59(1)& 0.251(2)\\
	& 0.5 &0.08982(4)& 0.159(1) & 0.275(2)& 1.59(1) & 0.251(1) \\[1ex]
\hline
\end{tabular}
\caption{Numerical estimate of critical points and critical exponents in BARW and 
PCPD for various values of $\epsilon$ obtained using Monte carlo simulations. The numbers
within  parenthesis represent the error in the last place of decimal. The errors 
are determined from eye estimate in fitting the exponents in the power laws and the scaling function. }
\label{table:exponents}
\end{table}

\begin{figure}
\centering\includegraphics[height=10cm]{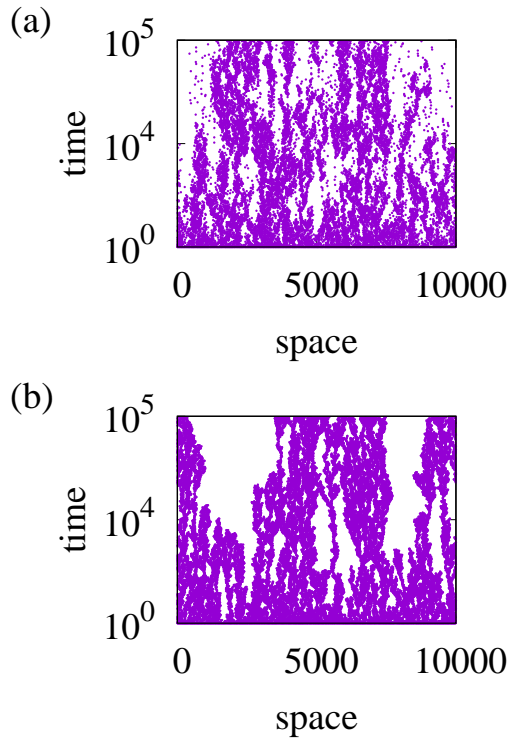}
\caption{space-time plots in PCPD for (a) $\epsilon=0$ and (b) $\epsilon=1/2$ at the critical points $q_c=$0.13353 and 0.08982 respectively.}
\label{fig:st_pcpd}
\end{figure}

\section{Results and discussion}
\par In BARW, for $m=1$ and $\epsilon=0$, our 
results match well with the DP 
exponents obtained previously \cite{pc_class}. For any $\epsilon>0$, the 
exponents remain same as that for $\epsilon=0$. Therefore, we conclude that 
the bias does not affect the DP universality class. 
The BARW model for $m=4$ belong to the PC 
class \cite{pc_class} in the absence of any bias $(\epsilon=0)$. Our results 
indicate the introduction of 
the bias drastically changes the exponents of the 
PC universality class. This is surprising 
because the bias does not the affect the parity of the system, 
neither creates any long-range interaction, nor does it gives rise 
to additional conservation laws. It would be interesting to 
know the universality class that the bias gives rise to for the PC class.  
\par Our simulations show that the PCPD class crosses over to DP for any 
nonzero $\epsilon$. A characteristic feature of PCPD 
is the survival of solitary diffusing  walkers for large times. When diffusion
 of single walkers is blocked in PCPD, its critical behavior 
is same as DP \cite{pcp_jensen}. The bias $\epsilon$ tends to supress the diffusion of 
single walkers away from its parent cluster. For $\epsilon=1/2$, a single walker cannot leave a cluster  and diffuse as a solitary walker. In Figure \ref{fig:st_pcpd}
we show the space-time plots for $\epsilon=0$ and $1/2$ in PCPD 
at the respective critical points. Clearly, the solitary diffusing walkers 
do not survive for large times for $\epsilon=1/2$ as compared 
to when $\epsilon=0$. In fact, this effect of solitary 
diffusing walkers not surviving for 
large times  seems to happen for smaller $\epsilon$ values also.  
This is possibly the reason 
that for any $\epsilon>0$, PCPD crosses over to DP. 
\par We conclude with the following observations. The universality 
class of DP is robust and remains unaffected by perturbations which 
alters the diffusion dynamics as long as the parity symmetry 
is unaltered. The same cannot be said for the PC and the PCPD
universality classes. Parity alone does not guarantee the PC critical behavior, and 
the PC class may change under  pertubations of the diffusion dynamics. But, as long as 
parity is kept unchanged, PC class does not seem to go to DP class. The 
PCPD critical behavior is rather unstable. Although it is not affected by parity, perturbations which arrests the diffusion of single walkers makes 
it cross over to DP.

\end{document}